# Optical absorbers based on strong interference in ultra-thin films

## Mikhail A. Kats[1,*] and Federico Capasso[2]

[1]Departments of Electrical and Computer Engineering, and (by courtesy) Materials Science and Engineering, and Physics, University of Wisconsin - Madison, Madison, Wisconsin 53706, USA

[2]School of Engineering and Applied Sciences, Harvard University, Cambridge, Massachusetts 02138, USA

* Corresponding author (mkats@wisc.edu)

Abstract:

Optical absorbers find uses in a wide array of applications across the electromagnetic spectrum, including photovoltaic and photochemical cells, photodetectors, optical filters, stealth technology, and thermal light sources. Recent efforts have sought to reduce the footprint of optical absorbers, conventionally based on graded structures or Fabry-Perot-type cavities, by using the emerging concepts of plasmonics, metamaterials, and metasurfaces. Unfortunately, these new absorber designs require patterning on subwavelength length scales, and are therefore impractical for many large-scale optical and optoelectronic devices.

In this article, we summarize recent progress in the development of optical absorbers based on lossy films with thicknesses significantly smaller than the incident optical wavelength. These structures have a small footprint and require no nanoscale patterning. We outline the theoretical foundation of these absorbers based on "ultra-thin-film interference", including the concepts of loss-induced phase shifts and critical coupling, and then present several applications, including ultra-thin color coatings, decorative photovoltaics, high-efficiency photochemical cells, and infrared scene generators.

## I.    Introduction

Numerous applications in optics and optoelectronics across the electromagnetic spectrum require efficient and compact optical absorbers. This necessity has spurred significant research activity into the development of thin electromagnetic absorbers from the ultraviolet (UV) to the radio frequency (RF) spectral ranges.

At optical frequencies, narrowband optical absorbers based on Fabry-Perot cavities have been used for resonant cavity enhanced (RCE) photodetectors and similar devices, where the field buildup in the cavity increases the percentage of light absorbed in the active region, albeit limiting the performance to a narrow wavelength range [1] [2]. Materials featuring electroabsorption have been incorporated into such resonant cavities to make voltage-driven reflection modulators [3] [4]. In the RF regime, absorbers such as Salisbury screens [5], in which a resistive sheet is placed a quarter wavelength above a ground plane, and related structures such as Jaumann absorbers [6] and resistively loaded high impedance surfaces [7] [8], have been used for minimizing the radar cross-section of objects for stealth technology [9]. For such applications, compact, broadband absorbers are desirable, and bounds have been derived on the relationship between absorber thickness and maximum achievable absorption bandwidth [10].

More recently, the rapidly increasing ubiquity of micro- and nanofabrication capabilities have led to the development of ultra-thin absorbers utilizing the emerging concepts of metamaterials and metasurfaces [11] [12]. These absorbers are typically significantly thinner than the wavelength, and consist of a metallic back-reflector, a thin dielectric spacer, and resonant structures on top. Such absorbers have been demonstrated in the RF, terahertz (THz) [13] [14], mid-infrared [15] [16], near-infrared [17] [18], and visible [19] [20] ranges. Other, more exotic approaches for absorbers are being explored as well, including the use of lossy anisotropic crystals [21].

All of the aforementioned absorber designs either (1) require nanoscale patterning which dramatically increases the fabrication costs and limits the maximum realistic area of such devices, or (2) are not very thin compared to the wavelength of light. There is significant benefit to realizing ultra-thin light-absorbing structures that require no micro- or nano-scale patterning. In this review, we focus on recent developments on such absorbers based on ultra-thin highly lossy films.

## II.   Absorption in an ultra-thin highly lossy film

The starting geometry we will examine is a simple three-material structure, in which a thin film of thickness $h$ and complex refractive index $n_2$ is sandwiched between air or another transparent dielectric with refractive index $n_1$ and a semi-infinite substrate with complex refractive index $n_3$ (Fig. 1). The field reflection and transmission coefficients for this structure can be found in many textbooks (e.g. refs. [22] [23] [24] [25]), and are sometimes referred to as the Fresnel-Airy formulas:

$$r = \frac{r_{12} + r_{23}e^{2i\beta}}{1 + r_{12}r_{23}e^{2i\beta}} \qquad (1)$$

$$t = \frac{t_{12}t_{23}e^{i\beta}}{1 + r_{12}r_{23}e^{2i\beta}} \qquad (2),$$

where $r_{ij}$ and $t_{ij}$ are the polarization-dependent Fresnel coefficients, $\beta = n_2 k_0 \cos(\theta_2)h$, $k_0 = \omega/c$ is the free-space wave number, $\omega$ is the angular frequency of the incident light, and $c$ is the speed of light in vacuum.

Assuming that all of the materials are isotropic and non-magnetic (i.e. permeability $\mu = \mu_0$), the field Fresnel reflection and transmission coefficients for s- and p- polarization from medium $i$ to medium $j$ are defined as

$$r_{ij}^s = \frac{n_i\cos(\theta_i) - n_j\cos(\theta_j)}{n_i\cos(\theta_i) + n_j\cos(\theta_j)} \qquad (3)$$

$$r_{ij}^p = \frac{n_j\cos(\theta_i) - n_i\cos(\theta_j)}{n_j\cos(\theta_i) + n_i\cos(\theta_j)} \qquad (4)$$

$$t_{ij}^s = \frac{2n_i\cos(\theta_i)}{n_i\cos(\theta_i) + n_j\cos(\theta_j)} \qquad (5)$$

$$t_{ij}^p = \frac{2n_i\cos(\theta_i)}{n_j\cos(\theta_i) + n_i\cos(\theta_j)} \qquad (6)$$

The angles $\theta_i$ are related to the incident angle $\theta_1$ by Snell's law, $\theta_i = \sin^{-1}\left(\frac{n_1}{n_i}\sin(\theta_1)\right)$, and can in general be complex-valued, which signifies loss in the material. The reflected power is then $R = |r|^2$, and the power transmitted through the interface can be written as $T_{s,int} = |t^2|\frac{Re[n_f\cos(\theta_f)]}{Re[n_i\cos(\theta_i)]}$ for the s-polarized case and $T_{p,int} = |t^2|\frac{Re[n_f\cos(\theta_f^*)]}{Re[n_i\cos(\theta_i^*)]}$ for the p-polarized case, where the subscripts $i$ and $f$ identify the initial and final materials (i.e. the superstrate and substrate, respectively), and the asterisk denotes the complex conjugate [26]. If the substrate is lossless, then $T_{(s \text{ or } p),int}$ is equal to $T$, the total power transmitted through the structure; conversely, for a lossy infinite substrate, $T = 0$ since the light is gradually absorbed as light propagates deeper into the substrate material.

Since a uniform film does not scatter light and energy must be conserved, the absorbance in the film $A_{film}$ is given by

$$A_{film} = 1 - R - T_{int} \qquad (7),$$

where the polarization subscripts are suppressed for brevity. If the substrate is transparent, then the total absorbance $A = A_{film}$; otherwise, some of the incident power can also be absorbed in the substrate, and $A = 1 - R$. For many applications, it is desirable that an absorber maximizes either $A_{film}$ or $A$, minimizes $R$, and is as thin as possible – ideally with a thickness much smaller than the wavelength of light.

Clearly, maximizing $A$ requires the presence of optical losses (i.e. a non-zero imaginary part of the refractive index, $n$) in the thin film, the substrate, or both. However, somewhat counter-intuitively, maximizing $Im[n]$ does not maximize the resultant absorption, because high losses can also result in the fields being efficiently reflected from the lossy material. This effect can be understood from the perspective of impedance; for non-magnetic materials, the wave impedance $Z$ is related by the refractive index as $Z = Z_0/n$, where $Z_0 \cong 377 \ \Omega$ is the impedance of free space [27]. Given a large imaginary part of $n$, the modulus of $Z$ becomes very small, leading to a large impedance mismatch between a wave in any lossless dielectric and in the lossy material. Therefore, a balance between field penetration and optical loss is required to maximize absorption.

## III.     Lossy film bounded by transparent dielectrics

This problem of maximizing absorption in a three-material thin film geometry has been studied for over 80 years. One result that has been derived several times in various contexts concerns the optimum refractive index of an ultra-thin lossy film with a transparent substrate and superstrate [28] [29] [30] [31]. Given a lossless substrate and superstrate with the same refractive index ($n_1 = n_3 = Re[n_1]$), there exists a combination of $h \ll \lambda$ and $n_2$ such that $A$ reaches a peak value of 0.5. This condition occurs when $Re[n_2] \cong Im[n_2] \cong \sqrt{(n_1 + n_3)\lambda_0/(4\pi d)}$ [30] (Fig. 1). For example, for $n_1 = n_3 = 1$, $\lambda_0 = 900$ nm, $h = 10$ nm, and normal incidence, the $A = 0.5$ condition is reached when $Re[n_2] \cong Im[n_2] \cong 3.8$ (Fig. 1(c)).

This condition is quite tolerant to small changes in both $n_2$ and $h$, as can be seen by the sparsely spaced contours in Fig. 1; for example, $A = 0.48$ for $n_2 \cong 5 + 4i$. As a result, this phenomenon can occur accidentally, such as in the case of failure of high-power microwave windows due to enhanced absorption in thin films of contaminants [29]. Note that for much larger values of $Re[n_2]$ or $h$, there exist additional maxima in absorption when $Im[n_2] \ll Re[n_2]$. These maxima correspond to coupling to Fabry-Perot modes and can have $A > 0.5$, but are no longer in the "ultra-thin-film" limit of $h \ll \lambda$.

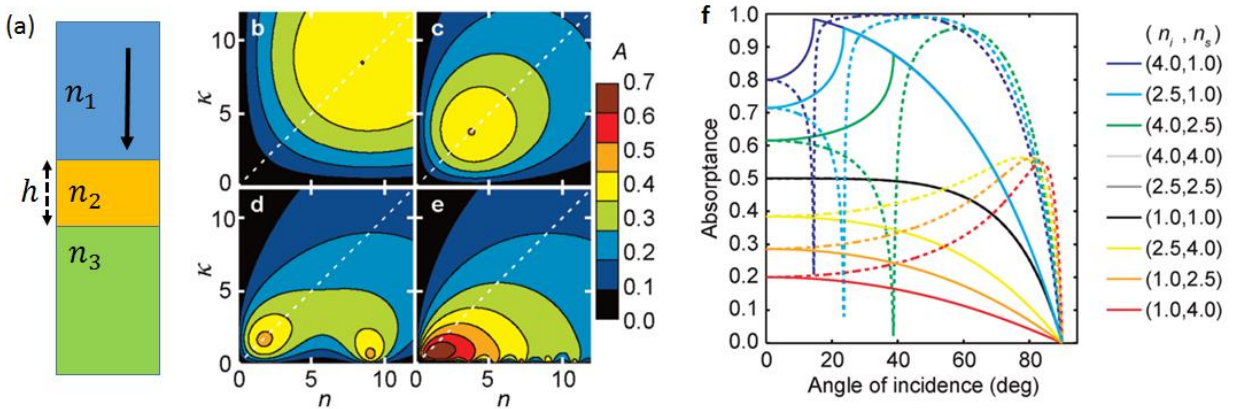

Figure 1. (a) Thin film with thickness $h$ and complex refractive index $n_2$, with light incident from a transparent superstrate with refractive index $n_1$, on a transparent substrate with index $n_3$. (b-e) Absorption maps of a thin film with $n_2 = n + ik$, with $n_1 = n_3 = 1$, and $h = 2$ nm (b), 10 nm (c), 50 nm (d), and 250 nm (e), respectively, at free-space wavelength $\lambda_0 = 900$ nm. (f) Absorbance vs. incident angle for p- (dashed) and s-polarization (solid). The dip for p-polarized light is at the critical angle between the film and the substrate. Adapted with permission from ref. [30]. Copyright (2010) American Chemical Society.

Interestingly, for film thickness much smaller than the wavelength, $A_{limit} = 0.5$ remains the maximum attainable absorption for any value of $n_1 = n_3$, as well as for any angle of incidence $\theta$ both for s- and p-polarization. Note, however, that for oblique incidence $Re[n_2] \cong Im[n_2]$ is no longer the condition for which $A$ reaches 0.5.

To surpass this absorption limit for ultra-thin films bounded by a transparent substrate and superstrate, the material from which the light enters must have a refractive index ($n_1$) larger than that of the substrate ($n_2$) [32] [33] [30]. For normal incidence, $A_{limit}$ can be shown to be $n_1/(n_1 + n_3)$, and it can be increased to 1 for oblique incidence and a precise selection of $n_2$ (Fig. 1(f)).

As shown in Fig. 1(f), $A$ can remain close to 1 over a significant range of incident angles (e.g. 20 ° to 60 °) for p-polarized incident light. This effect has been proposed for applications including ultra-thin superconducting single-photon detectors [32] and photovoltaics [30]. In applications like superconducting single-photon detectors [32], this type of light trapping mechanism is of profound importance because it offers the possibility of a large external quantum efficiency and allows the wires to be very thin, yielding large sensitivities. This configuration has been demonstrated by Driessen et al using a ~5 nm film of niobium nitride (NbN, $n_2 \cong 3.5 + 4.5i$ at $\lambda_0 = 775$ nm), a superconducting material [33]. Note that in refs. [30] and [32], the ultra-thin "highly absorbing film" was implemented not with a continuous film, but with two-dimensional arrays of oblate ellipsoids [30] and closely-spaced nanowires [32]. Given sufficiently small spacing between neighboring elements, these arrays can approximated as a continuous film. A similar mechanism has also been used to enhance the absorption in graphene, resulting in experimental broadband absorption values of ~0.1 in monolayer graphene in the visible [34], compared to the intrinsic value of 0.023 [35], and > 0.7 in multi-layer graphene [34]. For two-dimensional van der Waals materials like graphene, anisotropic forms of the Fresnel equations must sometimes be used, together with tensor forms of the refractive index [36] [37] [21].

## IV.    Lossy films on reflecting substrates

While the ultra-thin films in Fig. 1 can reach $A_{limit} = 1$, this condition is only satisfied for oblique incidence with a particular polarization, and only when the incident medium has a larger refractive index than the substrate. Furthermore, because the condition is reached at or beyond the critical angle, a prism or similar mechanism must be used to couple incident light into the structure. For many applications, these constraints are quite limiting, as an ideal absorber would not only be ultra-thin and continuous, but would also operate for both normal incidence and oblique incidence for both s- and p-polarizations, and would allow light coming from free space rather than from a high-index dielectric.

Over the last several years, such geometries have been realized using ultra-thin highly lossy films deposited on an opaque, reflecting substrate (e.g. refs. [38] [39] [40] [41] [42] [43] and others referenced later in the text). Intuitively, the use of a reflecting substrate rather than a transparent one should make it much easier to find a condition of high absorption. For a transparent substrate, one has to simultaneously minimize the reflectance $R$ and transmittance $T$ via interference effects, which poses a significant challenge, and is in many cases impossible. Conversely, if a substrate is opaque, $T = 0$ by definition, so maximizing $A$ requires simply minimizing $R$. Phrased as an optimization problem, using an opaque substrate halves the number of constraints. For this reason, the regions of high absorption in Fig. 1(f) occur at or beyond the critical angle, since total internal reflection suppresses $T$ without the need for any additional design.

An ultra-thin absorber with a reflecting substrate can be thought of as an absorbing anti-reflection (AR) coating. Therefore it is instructive to first consider the basic operating principle of the simplest AR coating, which minimizes $R$ and maximizes $T$ at normal incidence, given that both the substrate and superstrate are transparent: a quarter-wave film with $n_2 = \sqrt{n_1 n_3}$ and $h = \lambda_0/4n_2$ [22]. In this AR coating, there are no losses anywhere, and the film is designed such that reflection is suppressed via destructive interference. This can be verified using Eqn. (1), and can also be understood by decomposing the incident light into partial waves (Fig 2(a)), plotting the amplitude and phase of the reflected partial waves $r_0$, $r_1$, ..., in complex space (Fig. 2(b)). In this configuration, the initial reflected partial wave $r_0 = |r_0|e^{i\phi_0}$ has phase $\phi_0 = \pi$, whereas all of the other partial waves have phase $\phi_1 = \phi_2 = \cdots = 0$, leading to destructive interference due to the $\pi$ phase difference. Given a lossless film, substrate, and superstrate, $h = \lambda_0/4n_2$ is the thinnest film that yields a reflectance minimum.

A counter-intuitive situation occurs if the substrate in this geometry is replaced with a perfect electric conductor (PEC), which is an idealization of a metal with a very large plasma frequency [23]. Here all of the phases of the partial waves are identical to the case of the quarter-wave AR coating, seemingly implying that the reflected partial waves should destructively interfere due to the $\pi$ phase difference between $r_0$ and $r_{n>0}$. However, no destructive interference can ultimately occur, because there are no optical losses in the system and the substrate is opaque, so the reflectance $R = 1$. This configuration is known as a Gires-Tournois etalon, which is used as a phase-shifting element in certain optical systems [22] [38].

If a small amount of loss is present in the thin film ($0 < Im[n_2] \ll Re[n_2]$), and the substrate can be approximated as a PEC (i.e. $r_{23} \cong -1$), the quarter-wave condition does indeed correspond to destructive interference in the reflected waves, accompanied by a significant amount of optical absorption due to the field build-up in the film. The phasor diagram of such a structure is very similar to that of Fig. 2(b), with the key difference being that absorption in the film replaces the loss channel of out-coupling via the bottom interface (i.e. transmission). This effect can be thought of coupling to a resonance of the asymmetric Fabry-Perot (FP) cavity formed by the top and bottom interfaces, where the light is gradually absorbed as it circulates, just like in resonant-cavity enhanced (RCE) detectors and electro-optic modulators [1] [2] [3] [4]. The $R = 0$ (and hence $A = 1$) condition, when the reflected light is perfectly canceled, is an example of a phenomenon known as critical coupling, which occurs when the light is incident on a cavity at its resonance frequency, and the internal losses of the cavity are equal to the mirror losses [44] [39]. The concept of critical coupling can be invoked to explain a broad range of situations in electromagnetics where power is efficiently transferred to a resonator, such as in the case of plasmonic (or "metamaterial") perfect absorbers [45], and "coherent perfect absorbers" or "anti-lasers" [46].

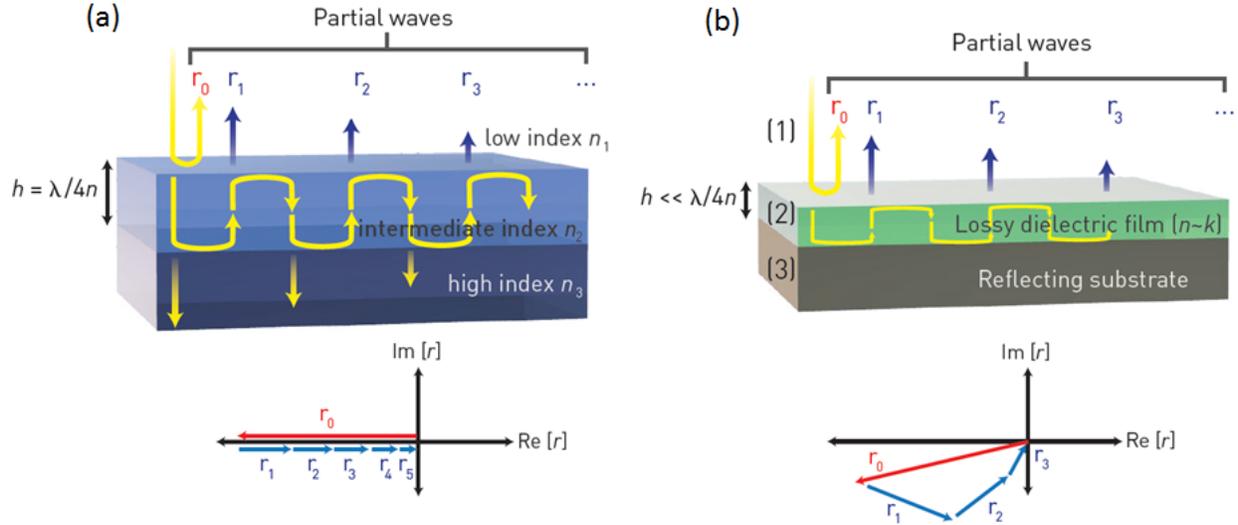

Fig. 2. Schematics and phasor diagrams for (a) a conventional quarter-wave anti-reflection coating, and (b) an absorber using an ultra-thin lossy layer on a reflecting substrate. Adapted from ref. [43], Phil Saunders Graphics / Source OPN, January 2014.

## V.     Perfect absorption in ultra-thin, highly lossy films

The remainder of this review will focus on the situation where this "lossy Gires-Turnois etalon" or "asymmetric Fabry-Perot" configuration is modified by introducing a large amount of optical loss in the thin film and/or substrate, which allows the absorber to become much thinner than the wavelength of light.

The zero-reflectance condition, which corresponds to "perfect absorption" when $T = 0$, can be obtained from the root of the numerator of Eqn. (1),

$$r_{12} + r_{23}e^{2i\beta} = 0 \qquad (8),$$

which corresponds to two equations which must be satisfied at the same time, an amplitude condition, and a phase condition:

$$|r_{21}| = |r_{23}|e^{-4\pi\frac{h}{\lambda_0}Im[n_2]} = |r_{23}|e^{-h\alpha_2} \qquad (9),$$

$$\phi_{23} + 2\phi_{prop} - \phi_{21} = 2\pi m \qquad (10),$$

where $\alpha_2$ is the absorption coefficient of the thin film, $r_{ij} = |r_{ij}|e^{i\phi_{ij}}$ so $\phi_{ij}$ are the reflection phase shifts, $\phi_{prop}$ is the propagation phase pick-up due to a single pass through the film, and $m$ is an integer. Note that $r_{ji} = -r_{ij}$.

Embedded in Eqn. (10) is an important subtlety: this is similar to, but not exactly, the phase relation that defines a resonator. In optics, a resonance condition corresponds to a situation where the complete round-trip phase accumulation is some multiple of $2\pi$ [22]. In fact, if the film has optical gain rather than loss (i.e. $\alpha_2 < 0$), it is straightforward to find the amplitude and phase conditions at which this geometry behaves as a laser at threshold (note: above threshold, lasing is a highly nonlinear phenomenon, so these equations no longer apply), which occurs when $R = |r|^2 \rightarrow \infty$:

$$1 = |r_{21}||r_{23}|e^{-h\alpha_2} \qquad (11),$$

$$\phi_{23} + 2\phi_{prop} + \phi_{21} = 2\pi m \qquad (12).$$

Evidently, the phase condition of an absorber (Eqn. 10) differs from that of a resonator on resonance (Eqn. 12) by the sign of the reflection phase contribution at the 2-1 interface, $\phi_{21}$. Making the low- but non-zero loss approximation above ($0 < Im[n_2] \ll Re[n_2]$), and assuming $n_1 < Re[n_2]$, as is the case for the vast majority of thin film dielectrics when the incident medium is air ($n_1 = 1$), $\phi_{21} \cong 0$, so the phase conditions in Eqns. (10) and (12) are equivalent. In this instance, it is fair to say that the $A = 1$ condition occurs on resonance. However, assuming that the loss in the thin film becomes very large ($Im[n_2] \gg 0$), the conditions are no longer equivalent, stressing the "critical coupling" terminology that is more easily applied to the low-loss case.

If the thin-film material is highly lossy, the first reflected partial wave $r_0 = r_{12}$ is complex-valued, and the phasor points away from the real axis in complex space Fig. 2(d). This enables a special condition for maximizing $A$ and minimizing $R$: when the film is highly lossy such that $Im[n_2]$ is close to $Re[n_2]$, and the substrate is reflective but far away from the PEC condition such that $|n_3|$ has the same order of magnitude as $|n_2|$, $R$ can be completely suppressed for film thicknesses $h \ll \lambda_0/4n$ even at normal incidence (Fig. 2(c, d)). In this case, the phasor addition (Fig. 2(d)) is less obvious than that of the quarter-wave AR coating (Fig. 2(b)), and traces out a loop in complex space. As long as the phasor sum is zero, total destructive interference for the reflected light is achieved, and $A$ reaches unity. The large loss is thus required to satisfy both the phase and amplitude conditions (Eqns. 9 and 10) in the case of an ultra-thin film.

This special set of requirements – that the film be very thin and very lossy, and the substrate be reflecting but not too reflecting – is less restrictive than it seems at first glance. Over the last several years, numerous material systems have been successfully used to achieve this condition across the ultraviolet, visible, infrared, and terahertz spectral ranges. Among these are semiconductors [e.g. germanium (Ge), silicon (Si), gallium arsenide (GaAs), iron oxide ($\alpha$-Fe$_2$O$_3$)] on metallic substrates [gold (Au), silver (Ag), aluminum (Al)] in the visible [38] [40] [41] [47] [48] [49], metal-dielectric composites on metals in the visible and near-infrared [50], phase-transition materials [e.g. vanadium dioxide (VO$_2$)] and semiconductors on polar dielectric substrates in their Reststrahlen band region (e.g. sapphire and silicon dioxide) in the mid-infrared [39] [51], and dielectrics on highly-doped semiconductors in the infrared [52]. The parameter space allowing for unity absorption in this thin-film configuration was explored in-depth by Park et al, showing that that to achieve high absorption in this geometry $Im[n_3]$, must be greater than 0.64 [47]. Park et al also clarified that the perfect absorption condition corresponds to the coupling of incident light to the Brewster mode supported by the thin-film structure [41].

In the remainder of this article, we will highlight several demonstrations of this special condition for various material systems, elaborate on the physics, and explore their applications.

## VI.    Coloring and filtering with ultra-thin layers

Some of the simplest demonstrations of this effect involve metal substrates and ultra-thin films of semiconductors at visible wavelengths. In an early example, films of amorphous Ge with thicknesses between 7 nm and 25 nm were deposited by electron beam evaporation on Au substrates (Fig. 3). Though Ge is an indirect-gap semiconductor, it is highly absorbing at visible frequencies due to direct electronic transitions that appear at high-enough photon energies [53], especially in its amorphous state (Fig. 3(a)). When the Ge was deposited on the Au substrates, a reflectance minimum appeared in the visible range for normal-incidence light, and the spectral position of this minimum red-shifted across the visible with increasing film thickness. The minimum occurs for thicknesses that are much smaller than the quarter-wavelength thickness typical of a conventional AR coating, even when the large refractive index of germanium is taken into account – e.g., at $\lambda_0 \cong 530$ nm for a 10 nm film. At this wavelength, the amorphous Ge refractive index is ~ $4.3 + 2.1i$ (Fig. 3(a)), resulting in $\phi_{21} \sim 11°$; the difference in phase between Eqns. 10 and 12, $2\phi_{21}$, corresponds to the phase accumulated via propagation through a thickness of about 7.5 nm.

While the $R = 0$ condition described by Eqns. (8-10) is not fully reached for the samples in Fig. 3(b), the minimum nevertheless corresponds to a very large absorption value, especially for such thin films. Calculations showed that the vast majority of the light that was not reflected was absorbed in the Ge layer rather than in the underlying Au (Fig. 3(c)), meaning that 70-80% of the incident light can be absorbed in semiconductor layers 10-15 nm thick. This result suggests potential optoelectronic applications, such as ultra-thin photovoltaics and photodetectors with high external quantum efficiencies.

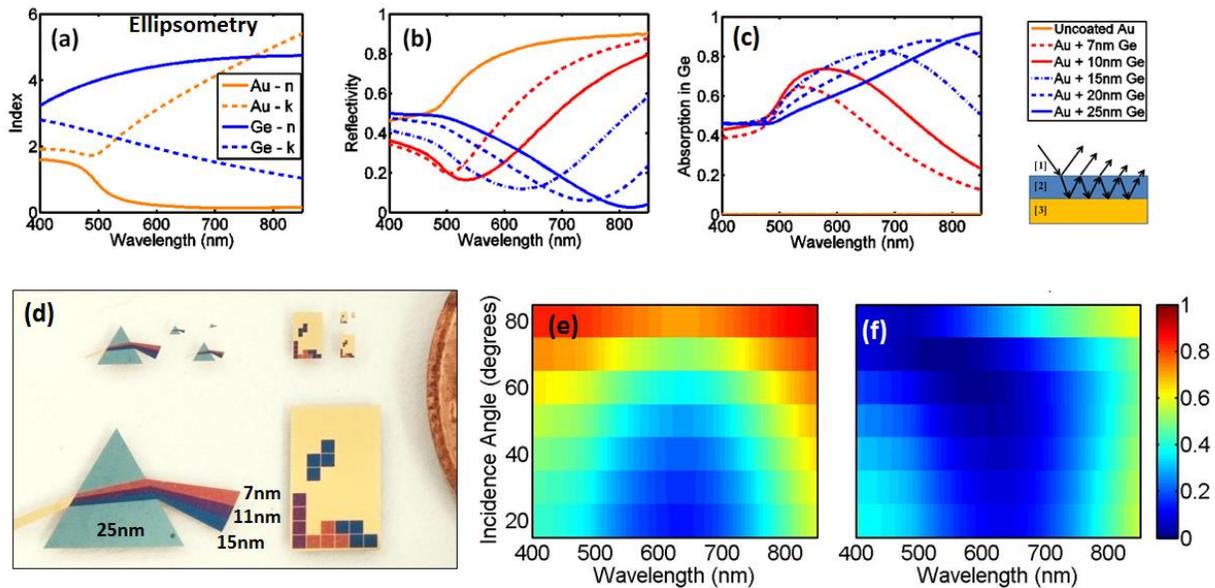

Fig. 3. Demonstration of an ultra-thin-film absorber geometry using Au (substrate) and amorphous Ge (film), for five different Ge thicknesses. (a) Measured complex refractive indices ($n + i\kappa$) of Au and Ge. (b) Measured reflectance at normal incidence. (c) Calculated fraction of incident light absorbed in the Ge layer. (d) Various color images formed by photolithography and repeated thin film depositions of a thick gold layer, and Ge films of 7, 11, 15, and 25 nm thickness. (e, f) Experimental color maps of reflectance as a function of wavelength and incident angle for a 15 nm Ge film on Au for s- (e) and p-polarized (f) light. Adapted from ref. [38].

Because the reflectance minimum can be shifted across the visible range by changing the film thickness, this approach can be used to color materials or create color images. In Fig. 3(d), photolithography and thin film deposition were used to create images, including a prism and a scene from Tetris. Optically thick Au films (100 nm) were deposited on a glass slide, followed by repeated depositions of nanometer-thick films of Ge with thicknesses of 7, 11, 15, and 25 nm, resulting in pink, purple, light blue, and dark blue colors, respectively. Thickness differences as small as a monolayer are observable as color changes to the naked eye, suggesting that this effect can be useful for sub-nanometer thickness monitoring for thin-film deposition systems [38]. The range and purity of accessible colors can also be substantially increased by adding additional films of subwavelength thickness [54] [55], which can simultaneously serve as capping layers to protect the color coating against chemical or physical erosion [54]. To calculate the $R$, $T$, and $A$ for these multi-layer films, a variety of textbook analytical methods can be employed, including the transfer matrix method [22] and the surface impedance method [25].

The deeply subwavelength film thickness also has the consequence of reducing the angle-dependence of the reflected spectrum. In most conventional examples of thin-film interference, such as soap bubbles and oil films on water, the emergent colors are highly angle-dependent (a.k.a "iridescent"), because the constructive and destructive interference conditions strongly depend on the angle-dependent phase difference that partial waves acquire as they propagate through the film. This is encoded in the $\theta$-dependence of $\beta$ in Eqns. (1, 2), which is proportional to the film thickness. Furthermore, the change in the propagation phase with respect to the incidence angle is almost completely canceled out by the angle-dependent reflection phase (Eqns. 3, 4) [48] [55], resulting in colors with minimal iridescence. For example, in the case of 15 nm Ge films, there is little change in the reflected spectrum between normal incidence and oblique incidence at a 45° angle for both s- and p-polarization (Fig. 3(e, f)). In fact, it has been shown that using exclusively optical methods (e.g. angle-dependent reflectance spectroscopy, ellipsometry, etc.), it is nearly impossible to determine if the colored objects such as those in Figs. 3 and 4 are colored by interference effects or by a smooth layer of conventional paint (see supplementary info in ref. [38]). This effect is closely related to a uniqueness problem in ellipsometry, where it is difficult to simultaneously determine the thickness and the refractive index of a film with thickness much less than the wavelength [56].

Interference colors similar to those of Fig. 3(d) and Fig. 4 can also be observed with transparent films on lossy reflective substrates. Examples include anodized titanium (Ti) and aluminum (Al), in which the top layer of the metal is electrochemically oxidized, forming thin films of transparent metal oxide [57]. Anodized metals can efficiently absorb light with oxide layer thicknesses corresponding to roughly 1/5[th] to 1/6[th] of the wavelength – thicker than the highly absorbing semiconductor coatings discussed in this review, but still thinner than the traditional quarter-wave AR coating [54]. The interference coloration that can be observed in anodized metals should not be confused with a process in which dyes are infused into the porous surface of an anodized materials and then thermally sealed, commonly used to color consumer products [58].

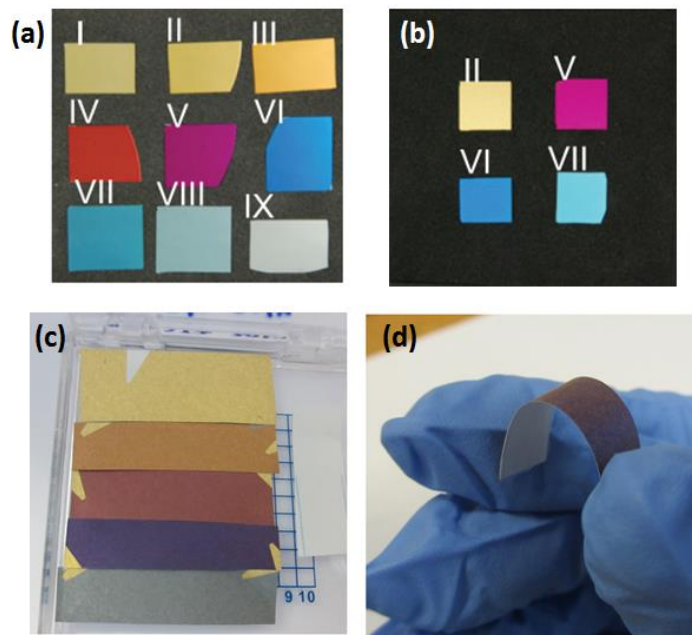

Fig. 4. (a, b) Interference colors generated by Si films of thickness from 0 to 30 nm (I-IX) on top of an opaque Au film, deposited on (a) a smooth, polished wafer, and (b) a rough, unpolished wafer. Reprinted with permission from ref. [49]. Copyright (2013), AIP Publishing LLC. (c, d) Interference colors generated by Ge films with thicknesses of 0, 7, 10, 15, and 100 nm on an opaque Au film, deposited on conventional cleanroom paper, which is a very rough and flexible substrate. Reprinted with permission from ref. [59]. Copyright (2014), AIP Publishing LLC.

The angle-insensitivity of color coatings made using ultra-thin lossy films on metal substrates means that they are tolerant of significant surface roughness, in contrast to thicker conventional interference coatings. For example, in refs. [38] and [49], interference colors from Si or Ge on Au substrates were shown to not change significantly even when the substrates became so rough as to be completely matte. In contrast, thin films deposited on insulating or semiconducting substrates exhibited colors that were angle-dependent [49]. The angle-insensitivity of lossy-film/metal coatings was also demonstrated using films deposited on conventional paper, which is a very rough and flexible substrate due to its intertwined cellulose fibers (Fig. 4(c, d)) [59]. The flexibility and matte appearance of the paper was not compromised by the coatings, indicating that they can be used to color almost any type of surface.

The ability to impart color to both smooth and rough surfaces without significant angle dependence makes this method a credible, albeit expensive, alternative to conventional paint. Whereas paints must be at least several microns thick (more typical is tens or hundreds of microns [60]), layers of metals and semiconductors totaling less than 50 nm can be used for coloration. In most instances, it is much more challenging to uniformly deposit large-area nanometer-thick coatings compared to conventional paint; however, for situations where weight is at a premium, e.g. space applications, this approach may be effective.

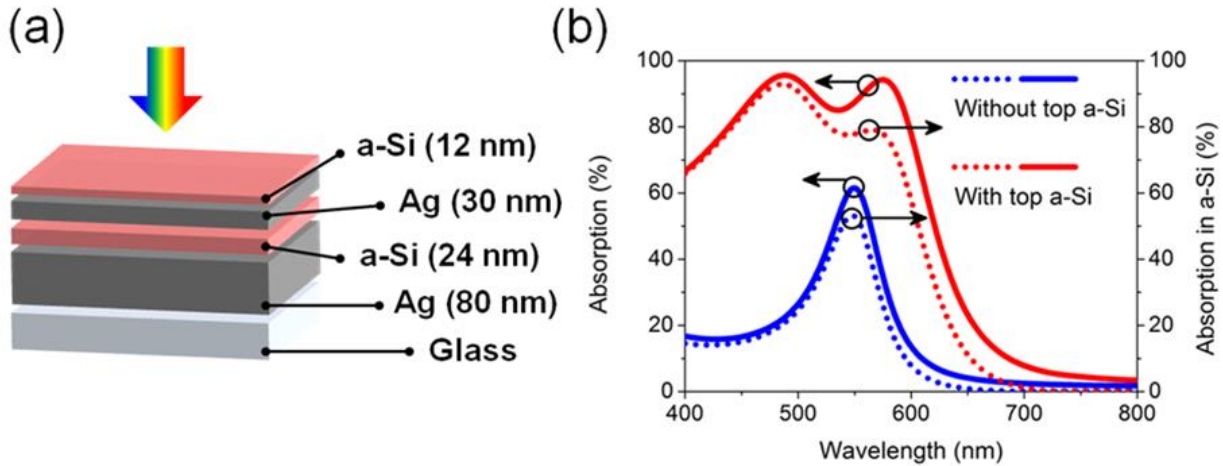

Fig. 5. (a) Schematic of a broadband absorber at visible frequencies using multiple layers of Ag and amorphous silicon (a-Si). (b) Simulated absorption spectrum without the top a-Si layer (blue) and with the layer (red). Reprinted with permission from ref. [61]. Copyright (2016), AIP Publishing LLC.

By incorporating additional ultra-thin layers of metals and dielectrics, the bandwidth of the absorption resonances can be substantially modulated. Decreasing the loss in the semiconductor (e.g. by using amorphous Si rather than Ge) can increase the quality factor of the Gires-Turnois-like cavity, and then the top interface reflectance can be increased to achieve the critical coupling condition by using a thin film of low-loss metal. This approach has led to the demonstration of reflection and transmission filters featuring high-purity colors with a degree of angle- and polarization-independence [48] [55]. By replacing the lossy semiconductor with a lossless dielectric like silica, and using the metal as the sole absorbing medium, the bandwidth of the absorber can be decreased even further [62]. Conversely, to increase the bandwidth of the absorber, multiple ultra-thin metal/semiconductor/metal cavities can be stacked (Fig. 5) [61] [63].

## VII.    Solar energy applications

The ability to enhance light absorption in ultra-thin semiconductor layers is of great significance for optoelectronic devices such as photodetectors, photovoltaic cells, and photoelectrochemical cells. In addition to potentially lowering the cost of devices via the reduction of required semiconductor material, ultra-thin layers that can absorb all or most of the incident light can overcome a fundamental trade-off between thickness and material purity that is related to carrier lifetimes in materials with defects [38] [64] [40].

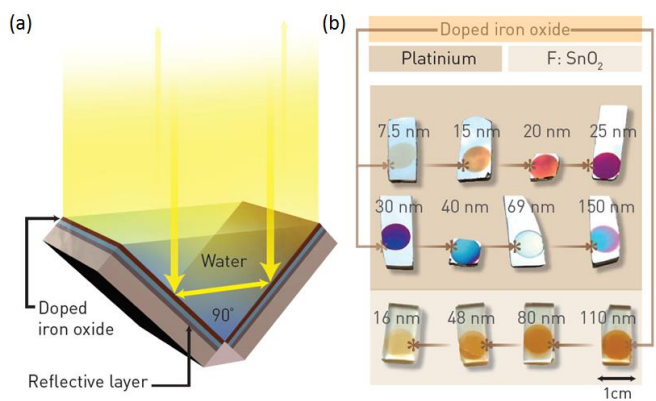

Fig. 6. (a) A water-splitting cell based on nanometer-thick Ti-doped iron oxide ($\alpha$-Fe$_2$O$_3$) layers on a metallized substrate. (b) Interference colors can be observed for $\alpha$-Fe$_2$O$_3$ films with thicknesses from 7 nm to 150 nm on platinum substrates, but not F:SnO$_2$, a transparent conductor. Adapted from ref. [40]. Phil Saunders Graphics / Source OPN, January 2014.

The first example of ultra-thin film absorption enhancement for optoelectronics was an ultra-thin water splitting solar cell based on iron oxide ($\alpha$-Fe$_2$O$_3$) films on metallic substrates (Fig. 6), demonstrated by Dotan et al [40]. $\alpha$-Fe$_2$O$_3$, a semiconductor with a band gap of 2.1 eV, is useful for photoelectrolysis because it is abundant, stable, non-toxic, and inexpensive. However, $\alpha$-Fe$_2$O$_3$ has poor transport properties; in particular, the short diffusion length of photo-generated holes means that any holes generated deeper than several tens of nanometers inside the $\alpha$-Fe$_2$O$_3$ layer will not make it to the surface with water, and are thus wasted [40] [65]. The authors investigated the optical absorption and water-splitting properties of Ti-doped $\alpha$-Fe$_2$O$_3$ layers with thicknesses from 10 to 100 nm, using various metallic substrates including Au, Ag, Al, and platinum (Pt), and found that these devices had much higher current densities than reference devices using the more-typical transparent conductor substrates. The results indicate that more than two thirds of the solar photons above the band gap of $\alpha$-Fe$_2$O$_3$ can be absorbed in sub-50 nm films, and most of the resulting photo-generated carriers can reach the film surface and react with water. This indicates a strong potential for this geometry for high-efficiency solar splitting cells.

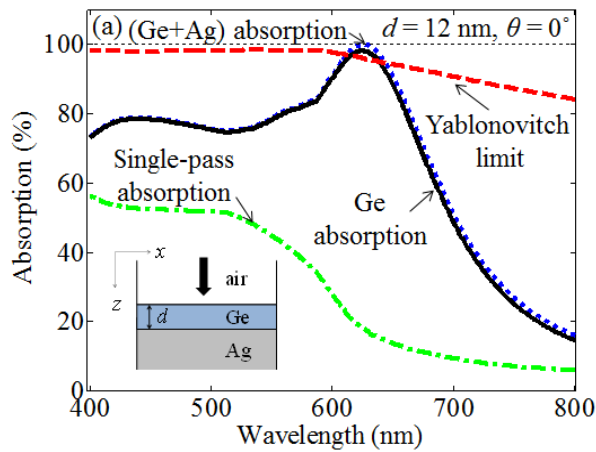

Fig. 7. Absorption in a 12 nm-thick Ge film on an Ag substrate compared to the single-pass absorption without the substrate, and to the Yablonovitch limit for light trapping in solar cells. Reprinted with permission from ref. [41]. Copyright (2014) American Chemical Society.

While the ultra-thin-film absorption mechanism has not yet been utilized to create photovoltaics with efficiencies that are competitive with state-of-the-art solar cells (e.g. 29% for single junction GaAs, 10% for amorphous Si (a-Si), and 11% for thin film organics [66]), several works have set up the foundation. For example, a study by Park et al showed that the enhanced absorption in ultra-thin films of semiconductors such as Ge, a-Si, GaAs, and copper indium gallium selenide (CIGS), can exceed the Yablonovitch limit, which describes the maximum achievable light absorption given the ray optics regime and materials with modest optical losses (Fig. 7) [41]. This limit is often used to gauge the effectiveness of light-trapping approaches, though it is not applicable in the cases of materials with large losses, or if special nanophotonic trapping mechanisms are used [67]. The ultra-thin-film mechanism also shows promise for organic photovoltaics (OPVs), which have thickness limitations due to exciton diffusion lengths that are on the order of 10 nm [68]. For example, a recent report showed that OPV films comprising P3HT:PCBM, a well-known OPV bulk heterostructure composite, can be made as thin as 60 nm and still absorb as much as 90% of the incident light when backed by an Al substrate [69], with similar results for other OPV materials [70].

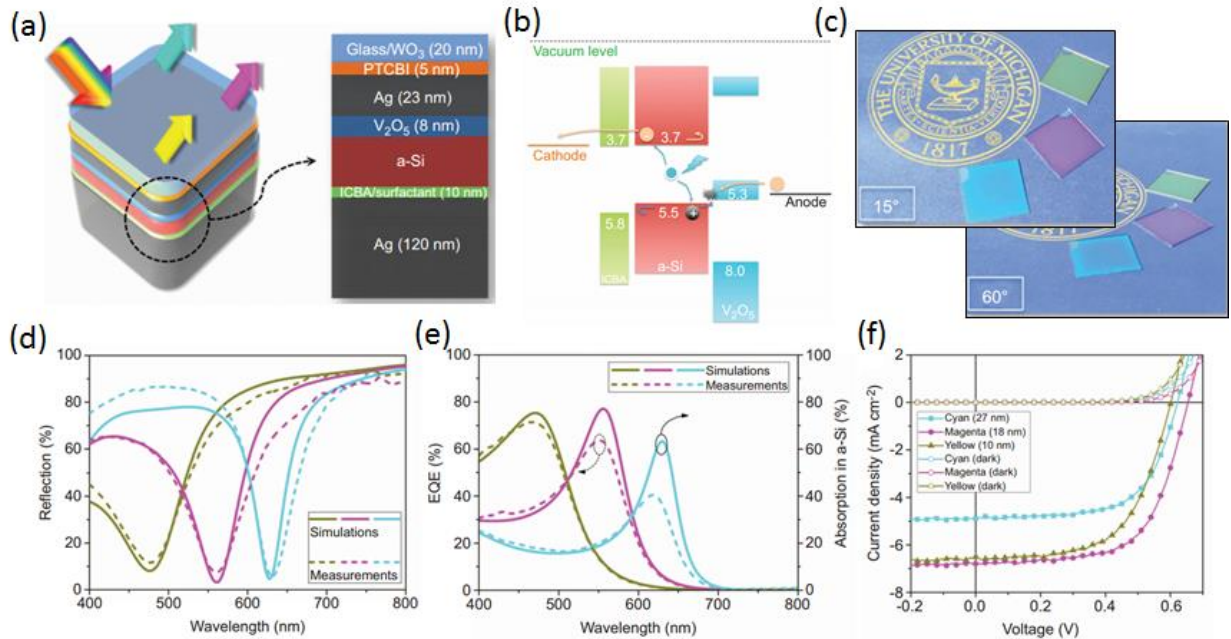

Fig. 8. Decorative ultra-thin photovoltaics. (a, b) Device structure and band diagram. The active layer is the a-Si film with thickness between 10 and 30 nm, bounded by an ICBA electron-transporting layer and a $V_2O_5$ hole-transporting layer. The thick bottom Ag layer is the cathode back-reflector, and the thin top Ag layer is the anode partial front-reflector. (c) Images of the devices with three different a-Si thicknesses yielding three different colors (10 nm, yellow-green; 18 nm, magenta; 27 nm, cyan), as seen at 15° and 60° angles of incidence. (d, e) Normal-incidence reflectance (d) and external quantum efficiency (EQE) (e) curves. (f) Current-voltage characteristics in the dark (empty symbols) and under AM1.5 illumination with irradiance of 100 mW/cm². Adapted with permission from Macmillan Publishers Ltd: ref. [71].

Though no overall efficiency records have been approached thus far, the ultra-thin film mechanism has been used to demonstrate a-Si photovoltaics that have large quantum efficiencies over relatively narrow wavelength ranges, and can have decorative qualities because of their structural colors in reflection [71] and transmission [42].

Typical a-Si solar cells use the traditional p-i-n heterojunction structure, have intrinsic layer thicknesses on the order of 100-300 nm [72], and efficiencies up to 10% [66], with carrier recombination in the p- and n-doped regions contributing to the low efficiency compared to crystalline Si cells. K. T. Lee et al utilized the ultra-thin-film absorption enhancement effect to design cells with a-Si thicknesses from 5 to 30 nm, and with the p- and n-doped Si regions replaced by a sub-10 nm film of vanadium oxide ($V_2O_5$), an electron-blocking and hole-transporting layer, and a thin organic film of indene-$C_{60}$ bisadduct (ICBA), an electron-transporting layer [42] [71] (Fig. 8). The back-reflector was an optically thick layer of Ag, used as the cathode, while the anode was a semi-transparent layer of Ag approximately 20 nm thick.

Because the active layer is much thinner than the typical carrier diffusion length in a-Si [73], the carrier recombination is negligible in these solar cells. External quantum efficiencies close to 80% have been measured (Fig. 8(e)), and track closely with the reflection spectrum (Fig. 8(d)). The overall power conversion efficiency in these cells was approximately 3%, compared to 10% for state of the art a-Si cells, which are an order of magnitude thicker. The reason for this modest overall efficiency is that the absorption peak is narrow-band, and does not reach 100% even at the optimum wavelength (Fig. 8(d)). This inefficiency might be overcome by using photon recycling approaches with cascaded cells with different active layer thicknesses [71] or, potentially, by using a geometry with multiple cavities such the one in Fig. 5 [61].

The metal/semiconductor/metal structure works as an optical filter in reflection [71], and therefore these solar cells can be made to display any number of interference colors (Fig. 8(c, d)), creating opportunities for decorative solar cells. Transmissive colors can be made in a similar way by using a semi-transparent metallic substrate [42].

VIII.    Infrared applications: static and dynamic absorbers and thermal emitters

The minimization of absorber thickness can have an even more profound significance at infrared frequencies compared to the visible range. In the visible, films with thicknesses on the order of the wavelength can be readily fabricated by conventional thin-film deposition techniques; in the infrared this can be a challenge. For example, given the standard long-wavelength infrared (LWIR) range, which covers roughly 8-15 $\mu$m [74], and typical IR materials which have refractive indices between approximately 1.5 and 4 [75], a quarter wavelength in the material corresponds to thicknesses between 0.5 and 2.5 $\mu$m. Depending on the deposition technique, thicknesses greater than ~ 1 $\mu$m can pose a challenge, so ultra-thin-film structures can be particularly beneficial.

However, at these long wavelengths, ultra-thin-film absorbers can no longer utilize conventional metals (e.g. Au, Ag, Al) as back-reflectors, because they are too close in behavior to PECs; i.e. $|n_3| \gg |n_2|$, so the reflection phase $\phi_{23}$ is always $\pi$, irrespective of the value of $n_2$. Fortunately, many materials exist which have complex refractive indices in certain regions of the infrared that are similar to those of conventional metals at higher frequencies, including polar dielectrics [39] [51] [76], highly-doped semiconductors [52] [77], and conducting oxides [78]. Highly doped semiconductors and conducting oxides can be thought of as dilute metals: they have a lower carrier concentration compared to conventional metals and, correspondingly, a lower plasma frequency, which is closer to the infrared range. The metal-like refractive index of polar dielectrics, including sapphire ($Al_2O_3$), silicon dioxide ($SiO_2$), silicon carbide (SiC), gallium nitride (GaN), and others, can be found in a portion of the infrared between the longitudinal and transverse optical phonon frequencies of the material. This region, referred to as the Reststrahlen band, can span a significant wavelength range; for example, for sapphire, it covers roughly 11 to 15 $\mu$m [79]. Materials with a Reststrahlen band at a desired wavelength can be found across the entire mid- and far-infrared spectral range, from 6 $\mu$m to 300 $\mu$m [76].

The first example of an infrared absorber based on ultra-thin planar films utilized sapphire as the "metal-like" substrate, and polycrystalline vanadium dioxide ($VO_2$) as the lossy layer [39]. $VO_2$ is a widely studied correlated-electron material that features an insulator-metal phase transition (IMT) that can be triggered thermally (at around 70 °C), electrically, optically, or via application of stress [80] [81]. Across the phase transition, $VO_2$ transitions from a semiconducting state (gap ~ 0.6 eV) to a metallic one. In the infrared, this corresponds to a transition from a low-loss dielectric, to a highly lossy dielectric in its intermediate state, to a lossy metal [82] [83]. The absorbing condition of Eqns. (8-10) can be reached when the $VO_2$ is in its lossy dielectric state.

The experimental demonstration is shown in Fig. 9 for a $VO_2$ film thickness of 180 nm. At low temperatures, the reflectance $R$ roughly follows the reflectance of sapphire, with the large value of $R$ in the 11-15 $\mu$m Reststrahlen band region. The film is much thinner than the wavelength across the entire measurement range, so no Fabry-Perot features are observed. At high temperatures, above the phase transition temperature, $R \sim 0.8$ across the entire range, a result of the Drude-like optical conductivity of the VO$_2$ in its metallic state. At an intermediate temperature of ~70 °C, however, $R$ drops to 0 for $\lambda_0 \cong 12\ \mu m$, a result of the perfect absorption condition of Eqns. (8-10). At this point, approximately 90% of the incident light is absorbed into the VO$_2$, while the remaining 10% absorbed in the top 1-2 $\mu m$ of the sapphire [39]. Just as with the ultra-thin absorbers in the visible in Sections VI and VII, this reflectivity minimum has low sensitivity to the incident angle, with $R < 0.01$ for angles between 0° and 30° for both s- and p- polarization (see suppl. info of ref. [39]).

Because of the large infrared absorption in an ultra-thin layer, this configuration is being considered for thermal detectors similar to bolometers, utilizing the phase transition to increase the detector sensitivity. While the thermally tunable absorber is unlikely to be useful for modulation applications due to the large thermal time constant, a similar geometry could be used if the phase transition is triggered electrically or optically, which has been shown to occur at picosecond time scales [80].

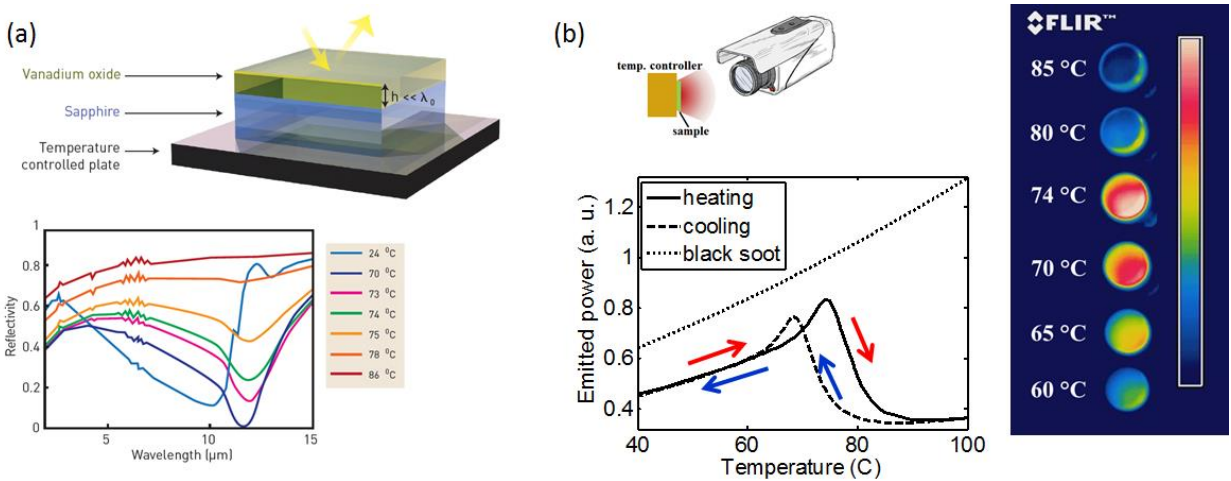

Fig. 9. (a) Temperature-dependent reflectance of a VO$_2$ film (thickness $h = 180$ nm) on a sapphire substrate, featuring perfect absorption around 70 °C and $\lambda_0 = 12\ \mu m$. Adapted from refs. [43] [39]. Phil Saunders Graphics / Source OPN, January 2014. (b) Thermal emission measurements on a sample similar to that of (a), demonstrating negative differential thermal emittance in the 75-85 °C range. The plot shows the measured thermally emitted power as a function of temperature. The hysteresis is intrinsic to the phase transition in VO$_2$. The right-most panel shows six snapshots of the device taken using a commercial LWIR infrared camera for increasing temperatures from 60 to 85 °C. Adapted from ref. [84].

As in the visible, a wide variety of lossy materials can be used as thin films for infrared absorbers. Phase transition materials like VO$_2$ are appealing because of their large inherent tunability, but these materials can be difficult to reliably grow, and they may not be compatible with certain established fabrication processes. Cleary et al investigated absorbers using highly doped Si films as the lossy layers on sapphire substrates, showing that $> 95\%$ absorption can be achieved in ~600 nm-thick Si films for a wavelength of ~12 $\mu m$ [51]. Streyer et al used thin films of undoped Ge (thicknesses between ~80 and 600 nm for $\lambda_0 \sim 5 - 15\ \mu m$), which is transparent in the infrared, on top of a highly doped Si substrate that is metal-like due to its large carrier density (Fig. 10) [52]. Since there are no losses in the film, all of the absorption takes place in the doped Si substrate, and this geometry can be viewed as the mid-infrared equivalent of the anodized metallic films in the visible [58], as described above in Section VI. Yen and Chung demonstrated a far-infrared ($\lambda_0 \cong 38 \ \mu m$) absorber using a 1.2 $\mu$m-thick GaAs film on an Au substrate [85], though, in this case, the reduction of physical thickness compared to the free space wavelength was a result of the large value of the refractive index ($Re[n_2] \sim 10$) rather than the "ultra-thin-film" interference effect.

One application that has emerged from research into optical absorbers is selective thermal emitters/radiators. Thermal radiation is the physical phenomenon responsible for the majority of light in the universe: every object not at absolute zero emits broadband thermal radiation, which peaks at a wavelength related to its temperature. In the past, thermal sources have often been thought of as being highly incoherent, emitting light indiscriminately over a broad range of wavelengths, angles, and polarizations. However, recent studies have shown that significant control over thermal emission can be obtained by using tailored absorbers.

The spectrum and intensity of thermal radiation emitted by any object at thermodynamic equilibrium is, in general, a function of its temperature and emissivity. The spectral irradiance $I(\lambda_0, T)$, a distribution function with units of power emitted per unit area, per unit wavelength, per unit solid angle (W/m$^3$/sr), can be written as

$$I(\lambda_0, T) = \frac{2hc^2}{(\lambda_0)^5} \left[ \frac{1}{e^{\frac{hc}{\lambda_0 k_B T}} - 1} \right] \epsilon(\lambda_0, \dots) \qquad (13),$$

where $c$ is the speed of light, $h$ is Planck's constant, $k_B$ is the Boltzmann constant, $T$ is the temperature in Kelvin, and $\epsilon(\lambda_0, \dots)$ is the emissivity [86]. The emissivity is wavelength-dependent, and can in, principle, depend on other parameters as well. When $\epsilon(\lambda_0, \dots) = 1$, Eqn. (13) is the well-known Planck's law, which describes thermal emission by a blackbody. A thermodynamic relationship known as Kirchhoff's law of thermal radiation relates $\epsilon(\lambda_0, \dots)$ to the absorptivity $A$ of any object:

$$\epsilon(\lambda_0, \dots) = A \qquad (14).$$

The ellipsis (…) indicates that, in general, both $A$ and $\epsilon$ can be dependent on any number of variables, including wavelength, angle, temperature, applied field, etc. When integrated over all wavelengths and all angles within a hemisphere, the total power emitted per unit area (W/m$^2$) is given by the Stefan-Boltzmann law:

$$I(T) = \mathrm{E}(\dots)\sigma T^4 \qquad (15),$$

where $\sigma$ is the Stefan-Boltzmann constant, and $\mathrm{E}(\dots)$ is the integrated emissivity.

Eqn. (14) connects the problem of engineering thermal emission to the problem of engineering the optical absorption. Indeed, in the literature, engineering the wavelength- and/or angle-dependence of $\epsilon$ has been achieved using a variety of optical methods including surface texturing with disordered structures [87], gratings [88], photonic crystals [89] [90], optical cavities [91], as well as plasmonic and metamaterial absorbers [13] [92]. Tailored thermal emitters have a wide variety of applications, including infrared scene generation, thermal signature masking [84], and wavelength-selective thermal light sources [92], which are especially useful for thermo-photovoltaics [93].

The tunable and/or tailorable absorbers in Fig. 9(a) and 10(a) can be utilized to obtain a significant amount of control of thermal emission. For example, by adjusting the thickness of Ge layers on highly doped Si substrates, Streyer et al demonstrated that the spectrum of thermal emission can be controlled as a function of position on the sample (Fig. 10(a)) [52]. The inset of Fig. 10(b) shows an image of a thermally emitted logo that was recorded with an infrared camera when the sample was heated.

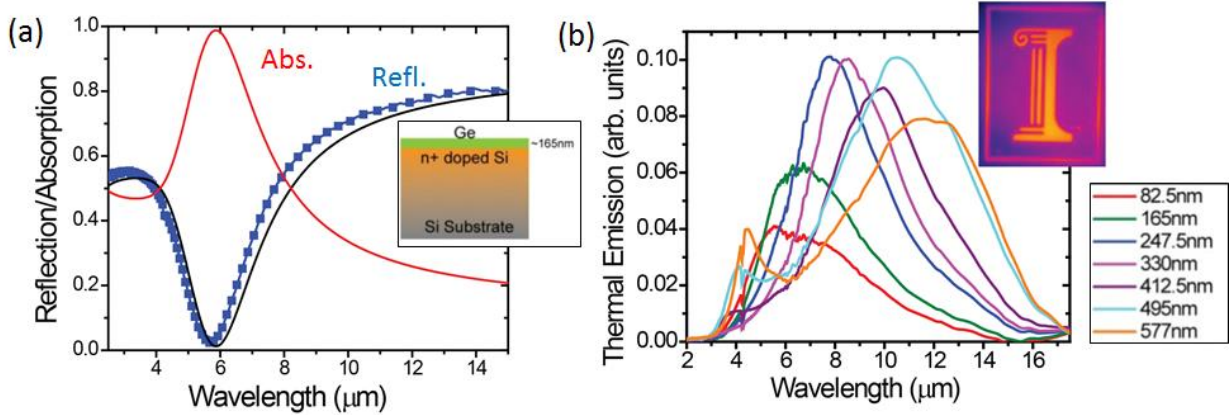

Fig. 10. (a) Experimental (squares) and calculated (lines) reflectance and absorbance of a 165 nm-thick film of Ge on a highly doped Si substrate, showing perfect absorption at ~ 6 $\mu$m. (b) Ge films with thicknesses from 80 to 600 nm result in absorbance/emissivity peaks across the mid-infrared. Shown is the emission measurement at $T = 450$ K. Inset: thermal emission image of a logo from a patterned 425 nm Ge layer. Adapted from ref. [52].

An intriguing aspect of Eqns. (13) and (14) is the potential dependence of the emissivity on parameters other than the wavelength. In particular, the VO$_2$/sapphire thin-film structure in Fig. 10(a) has a temperature-dependent absorptivity $A(\lambda_0, T)$ that varies dramatically and non-monotonically within the $70° - 80°$ range because of the VO$_2$ phase transition. As a result, both the emissivity $\epsilon(\lambda_0, T)$ and integrated emissivity E($T$) acquire a strong temperature dependence, and the expressions in Eqns. (13) and (15) can become non-monotonic with temperature. This is demonstrated in Fig. 9(b): the power emitted by the VO$_2$/sapphire structure in the $8 - 14$ $\mu$m atmospheric transparency band was measured as a function of temperature, observing a ~ 10 ° range of negative differential thermal emittance – a region where an increase in temperature led to a decrease in emitted power. This unique phenomenon is possible because $\partial \epsilon / \partial$T is very large (and negative) in the vicinity of the phase transition, overcoming the temperature dependence of the blackbody contribution. The effect can be confirmed by imaging the sample with a standard long-wave infrared camera (Fig. 9(c)), and observing that in the $75 - 85$ °C range, the sample appears cooler as the temperature is increasing. This temperature range be engineered by using other phase transition materials [80] or modifying the materials via doping [94] and defect engineering [95]. The continued development of thermal emitters with temperature-dependent emissivity may have significant applications in infrared camouflage, labeling and identification, and passive thermal regulation.

## IX.    Conclusion and Outlook

Due to their vast utility in optics and optoelectronics from the ultraviolet to the far infrared, engineered optical absorbers are a focus of numerous ongoing research and development efforts. Here we reviewed a small subset of these efforts, describing the theory, experimental demonstrations, and applications of ultra-thin optical absorbers based on lossy films. These absorbers typically have thicknesses in the 10-100 nanometer range, satisfying the rigorous size and weight requirements of modern micro- and nano-scale devices, and can be inexpensively fabricated on a large scale. We described the sometimes-counterintuitive physical concepts that govern ultra-thin-film absorbers, including loss-induced phase shifts and critical coupling, emphasizing the difference between the conditions of maximal absorbance and resonance.

The enhancement of absorption in ultra-thin layers has important consequences for numerous applications in photodetection, solar energy, filtering and coloration, and light emission. Decreasing the active region thickness of optoelectronic devices such as detectors, photodiodes, and solar cells, can reduce materials costs and simultaneously improve figures of merit like quantum efficiency. Even non-device applications such as color printing, structural color, and filtering can benefit significantly from the reduced overall weight and material use, enabled by the ultra-thin-film interference mechanism. The lack of angle sensitivity and ease of integration with flexible substrates makes these concepts widely applicable. We anticipate that as this field matures, applications utilizing the ultra-thin-film absorption mechanism will begin to augment and supplant existing technologies.

Acknowledgements

MK thanks J. Choy for a critical reading of the manuscript, and acknowledges startup funds from UW Madison. MK and FC both acknowledge the contributions of many current and former colleagues and collaborators, including R. Blanchard, P. Genevet, S. Ramanathan, Y. Zhou, M. Qazilbash, D. Basov, S. Zhang, D. Sharma, J. Lin, Z. Yang, Y. Cui, C. Ko, C. Ronning, and J. Rensberg. We also thank L. J. Guo, M. Brongersma, J. Park, F. Schlich, R. Spolenak, D. Wasserman, and C. Hagglund for contributing high-quality figures to this manuscript.

Key words

Thin-film interference, optical coatings, color coatings, structural color, optical absorbers, perfect absorbers, thermal emitters, optical filters, photovoltaics

Short bios

Mikhail Kats is an Assistant Professor of Electrical and Computer Engineering, as well as Materials Science and Engineering, and Physics (by courtesy) at the University of Wisconsin – Madison. He received his BS in Engineering Physics from Cornell University in 2008 and PhD in Applied Physics from Harvard University in 2014, working with Federico Capasso. During his graduate studies, he was the recipient of the National Science Foundation (NSF) Graduate Research Fellowship, and the Harvard Graduate Society Merit Fellowship. Since starting his position at UW-Madison in 2015, he received the Office of Naval Research (ONR) Young Investigator Award, and was selected for the Forbes "30 under 30" list in the Science category. Kats's work spans the fields of photonics, plasmonics, nanoscale science, and device physics, with contributions including the co-invention of optical metasurfaces based on phase discontinuities, and the discovery of strong thin film interference effects in ultra-thin highly absorbing films.

Federico Capasso is the Robert Wallace Professor of Applied Physics at Harvard University, which he joined in 2003 after 27 years at Bell Labs where his career advanced from postdoctoral fellow to Vice President for Physical Research. He has contributed to optics and photonics, solid-state electronics, nanotechnology, mesoscopic physics and QED phenomena such as the Casimir effect. Highlights of his research are band-structure engineering of new artificially structured semiconductors and devices, including the invention of the quantum cascade laser, superlattice avalanche photodiodes and resonant tunneling transistors; the first measurement of the repulsive Casimir force and in recent years pioneering work with his group on metasurfaces including the generalized Snell law and planar low aberration metalenses. He is the author of over 500 publications and holds 65 US patents. He is a member of the National Academy of Sciences, the National Academy of Engineering, the American Academy of Arts and Sciences (AAAS), the Academia Europaea and a foreign member of the Accademia dei Lincei (Italy). His awards include the IEEE Edison Medal, the American Physical Society Schawlow Prize, the King Faisal Prize, the SPIE Gold

Medal, the AAAS Rumford Prize, the IEEE Sarnoff Award, the Optical Society Wood Prize, the Franklin Institute Wetherill Medal, the Materials Research Society Medal, the Jan Czochralski Award for lifetime achievements in Materials Science and the Gold Medal of the President of Italy for meritorious achievement in science.